\documentclass[prd, twocolumn, nofootinbib]{revtex4}

\usepackage{amsmath}
\usepackage{amssymb}
\usepackage{latexsym}
\usepackage{graphicx,subfigure,graphics,epsfig,amsmath,amssymb}
\usepackage{hyperref}
\usepackage{bm}

\newcommand{\beq}{\begin{equation}}
\newcommand{\eeq}{\end{equation}}
\newcommand{\beqa}{\begin{eqnarray}}
\newcommand{\eeqa}{\end{eqnarray}}

\newcommand{\vp}{V_\phi}

\newcommand{\ophi}{\Omega_\phi}
\newcommand{\lam}{\lambda} 
\newcommand{\Lam}{\Lambda}

\begin{document}

\title{Old Dark Energy}

\author{Marina Cort\^{e}s${}^1$ and Eric V.\ Linder${}^{1,2,3}$} 
\affiliation{${}^1$Berkeley Lab, Berkeley, CA 94720, USA}
\affiliation{${}^2$University of California, Berkeley, CA 94720, USA}
\affiliation{${}^3$Institute for the Early Universe, Ewha Womans 
University, Seoul, South Korea} 

\date{\today}

\begin{abstract} 
Dark energy dynamics in the recent universe is 
influenced by its evolution through the long, matter dominated expansion 
history.  A particular dynamical property, the flow variable, remains 
constant in several classes of scalar field models as long as matter 
dominates; the dark energy is only free to diverge in behavior at 
recent times.  This gives natural initial conditions for Monte Carlo 
studies of dark energy dynamics.  We propose a parametrization for the 
later evolution that 
covers a wide range of possible behaviors, is tractable in making 
predictions, and can be constrained by observations.  We compare the 
approach to directly parametrizing the potential, which does not take 
into account the maturity of the dark energy dynamics. 
\end{abstract}


\maketitle

\section{Introduction \label{sec:intro}} 

Very little is known about the dynamics of dark energy, other than that 
at redshifts $z\gtrsim1$ the expansion history gives way to a matter dominated 
era.  However, this fact is powerful in narrowing the possible behaviors 
of dark energy because the matter-dominated Hubble friction plays a 
major role in the dark energy dynamics.  That is, the dark energy 
field $\phi$ evolves for a long time in a particular 
environment -- by the time dark energy comes to contribute significantly to 
the expansion it is an old, or  ``mature'', field. 

The influence of matter domination on the dark energy dynamics is very 
different than the slow-roll conditions for an inflationary field in 
the early universe.  Dark energy does not satisfy slow-roll conditions 
on the potential $V$ for most of its evolution, unless it is highly fine 
tuned \cite{paths}.  Instead, the matter domination can create a definite 
relation between the deviation of the dark energy density $\ophi$ 
from 0, the deviation of its equation 
of state $w$ from $-1$, and the characteristic scale of the 
potential, $(1/V)dV/d\phi$ (which is sometimes called the first slow-roll 
parameter, but here does not need to be small).  A particular 
combination of them is called the flow parameter by \cite{cahndl} 
and is essentially constant during matter domination. 

In this paper we examine the dark energy evolution after matter 
domination wanes, by parametrizing the deviation of the flow parameter 
from its constant value during matter domination.  Since the constancy holds 
up until quite late redshifts, $z\approx2$, the dark energy has a 
restricted ability to exhibit diverse dynamics by the present.  This 
allows the three parameter flow form we present to reasonably approximate 
a wide range of the possible behaviors. 

Any form containing few parameters will not in general be able to 
describe every 
possible behavior so interesting questions include whether it captures 
key physics, reasonably describes the major classes of behavior, and 
can be constrained by observations.  In particular, the dynamics should 
include the thawing and freezing classes \cite{caldlin}, which 
respectively depart from and approach cosmological constant behavior 
within specific regions of the $w$-$w'$ phase space.  But it should 
also be flexible enough to permit models outside these regions and have 
the physics and data determine which are viable. 

In \S\ref{sec:init} we discuss the implications of a long matter 
dominated era in guiding the dark energy dynamics and the role of the 
flow parameter.  We present our approach to evolution as dark energy 
increases in importance in \S\ref{sec:fform} and study the resulting 
behaviors.  Constraints on the $w$-$w'$ and flow parameters phase spaces 
from future observations are investigated in \S\ref{sec:condyn}.  We 
compare the flow approach to parametrizing the potential 
in \S\ref{sec:compare}, discussing the different weights these impose 
on the dynamics and the thawing vs.\ freezing classes.

\section{Mature Dynamics \label{sec:init}}

During the early universe accelerated expansion of inflation, the 
inflaton scalar field dominates the total energy density and so the 
field evolution is essentially wholly determined by its potential. 
Once inflation ends, the cosmic expansion is dominated by radiation 
and then matter.  The expansion rate -- the Hubble parameter $H$ -- 
is rapid and the Hubble drag influences the evolution of any light 
scalar field that may eventually cause a more recent universe acceleration. 

The evolution of this scalar field is determined by the Klein-Gordon 
equation 
\beq 
\ddot\phi+3H(\rho_b,\rho_\phi)\,\dot\phi+\vp=0\,, \label{eq:kg} 
\eeq 
where we have explicitly indicated the dependence of the Hubble parameter 
on the background energy density $\rho_b$ and the scalar field density 
$\rho_\phi$.  The energy density $\rho_\phi=V(\phi)+(1/2)\dot\phi^2$, 
the sum of the potential and kinetic energies, so we cannot determine 
the field evolution without knowing $V(\phi)$ and the initial conditions 
$\phi_i$ and $\dot\phi_i$.  Formally, then, we cannot say anything 
completely general about the dark energy dynamics. 

However, an initial kinetic energy much larger than the potential, 
$K\gg V$, will rapidly redshift away, $\dot\phi\sim a^{-3}$ as can be 
seen by neglecting the last term in Eq.~(\ref{eq:kg}).  
So we expect the kinetic energy of the field does not dominate, but 
rather the background fluid (radiation or matter) does. 
See \cite{ratrap,wett88,frie95} 
for early papers on cosmological scalar fields, or quintessence. 
The distance $\Delta\phi$ the field rolls can be estimated by 
approximating $\dot\phi\approx\Delta\phi/\Delta t$ 
and using $K=(1/2)\rho_\phi\,(1+w)$, where the equation of state 
$w=[(K/V)-1)]/[(K/V)+1]$.  Then 
\beq 
\Delta\phi\sim\sqrt{(1+w)\,\rho_\phi\,(\Delta t)^{2}}\,. \label{eq:dphik} 
\eeq 
If the Hubble friction is large, then the characteristic timescale 
of the field evolution may be the Hubble time $H^{-1}$, so the square root 
contains $\rho_\phi/H^2\sim\Omega_\phi$ which is small in the 
matter (or radiation) dominated era.  So the era of background 
domination has significant effect on the dark energy dynamics.  

If the potential dominates the kinetic energy then $1+w$ is very 
small as well and so both factors lead to $\Delta\phi\ll1$ during 
the background domination (although by today, with dark energy domination, 
the condition $\Delta\phi\ll1$ generally does not hold; see 
Sec.~\ref{sec:compare}).  
This class is known as thawing fields.  In other cases $1+w$ does 
not start small and the field can roll further. 

If the potential has particular forms, the field can possess an 
attractor trajectory (see, e.g., \cite{ratrap,wett88,ferjoy,zws,lidsch,swz}). 
In this case the equation of state is determined 
by the background component and the parameters of the potential, 
independent of initial conditions.  The potential and kinetic energies 
are locked in a set ratio such that the equation of state is constant 
during background domination; the fields are then known as trackers. 

In any case, the dark energy density is not exactly zero and the 
equation of state is not exactly $-1$.  It is interesting to consider 
how these deviations are related to each other in the background 
dominated era. 
Ref.~\cite{cahndl} showed (also see \cite{scherrersen}) that one could 
define a combination of these deviations, together with the 
characteristic scale of the potential, that held constant during the 
long matter dominated expansion.  Ref.~\cite{cahndl} defined 
\beq 
F\equiv\frac{1+w}{\ophi\lam^2}\,, \label{eq:fdef}
\eeq 
where $1+w$ is the equation of state deviation, $\ophi$ the dark energy 
density in units of the critical density, and $\lam=-(1/V)dV/d\phi$ 
gives a characteristic field scale.  Despite all these quantities varying 
as the universe expands and the field evolves, the flow combination $F$ 
remains almost constant. 

Written in terms of the dark energy dynamics, 
\beq 
w'=-3(1-w^2)\left[1-\frac{1}{\sqrt{3F}}\right]\,, \label{eq:wpf} 
\eeq 
where $w'=dw/d\ln a$.  Thus the long matter dominated era, which 
imposes $F=\,$constant until quite late times, $z\approx2$, (except for 
specifically fine tuned fields) effectively imposes constraints on 
the dark energy dynamics.  That is, the dark energy is an old field 
that lived for many $e$-folds of expansion in a matter dominated environment.  
Choosing an arbitrary dynamics amounts to either ignoring this physical 
influence or embracing a fine tuning of the field's initial conditions. 

A direct parametrization of the potential (see, e.g., 
\cite{huttur99,simon,liddlesahlen,sahlen2,liddle_clemson}), 
and Monte Carlo variation of its parameters {\`a} la the interesting 
dynamics work of \cite{hutpeir,sahlen2}, will not in general take into 
account the maturity.  While one could parametrize the equation of 
state in such a way as to incorporate the appropriate condition (see, 
e.g., \cite{caldlin,scherrersen,crittenden,paths}), we choose to study the 
flow parameter since it possesses the interesting properties of being 
nearly constant during matter domination and incorporating $\ophi$, the 
natural quantity for measuring the deviation from matter domination as it 
begins to break.

\section{Evolving Dynamics \label{sec:fform}} 

During matter domination, fields which slowly relax from being frozen 
by the high Hubble friction -- thawing fields -- possess $F=4/27$ 
\cite{cahndl}.  Equivalently, they thaw from the cosmological constant 
state along the trajectory $w'=3(1+w)$ according to 
Eq.~(\ref{eq:wpf}).  Tracker fields possess $F=1/3$ or $w'=0$.  Fields 
with other behaviors are certainly possible, but these behaviors will 
be sensitive to the specific initial conditions.  However in this 
section we will briefly consider other high redshift values for $F$. 

We are most interested in how the dark energy evolves as matter 
domination wanes, and as eventually dark energy comes to dominate by the 
present.  That is, we want to extend the asymptotic, matter dominated 
behavior to greater dynamical freedom in the present.  Calculating the 
first order (in the dark energy density fraction) corrections to the 
dynamics \cite{cahndl} gives a flow parameter behavior 
\beq 
F\approx F_0+\beta\,a^B\,, \label{eq:f0b} 
\eeq 
where $F_0$ is the matter dominated constant value, $\beta$ is related 
to the dark energy density $\ophi$, and $B=3$ (or $-3w_\infty$ where 
$w(z\gg1)=w_\infty$ in the thawing (or tracking) case. 

However, since today the ratio of dark energy density to matter density 
is nearly 3, we do not expect the first order solution to be valid. 
Numerical solutions, for example, show that $F$ can grow by factors of 
order two to three.  Therefore, although we will keep the form 
(\ref{eq:f0b}) we take $\beta$ and $B$ to be free parameters 
representing effective fits over the expansion history between high 
redshift matter domination and today.  

The form (\ref{eq:f0b}) has the advantage that, upon substitution into 
Eq.~(\ref{eq:wpf}), one can analytically solve for the dynamics $w(a)$. 
The result is 
\beqa 
w&=&\frac{\chi-1}{\chi+1}\,, \\ 
\chi&=&\frac{1+w_i}{1-w_i}\,\left(\frac{a}{a_i}\right)^{-6} \\ 
&\quad&\times \left[\frac{\sqrt{F}-\sqrt{F_0}}{\sqrt{F}+\sqrt{F_0}} \frac{\sqrt{F_i}+\sqrt{F_0}}{\sqrt{F_i}-\sqrt{F_0}}\right]^{6/(B\sqrt{3F_0})} \,, 
\eeqa 
where $F_i=F(a_i)$, $w_i=w(a_i)$, and $a_i$ is some initial scale 
factor, e.g.\ in the matter dominated regime.  The four parameters are 
therefore $F_0$, $\beta$, $B$, and $w_i$, and we choose $a_i=0.1$ (the 
exact value of $a_i$ is unimportant as long as it is well within the 
matter dominated era). 

As the flow $F$ increases as the dark energy density becomes more 
important, the field evolves further in the direction of the 
skating curve, $w'=-3(1-w^2)$, where the potential slope is unimportant 
and field moves -- skates \cite{liddlesahlen,lincurv} -- with a rapidly 
redshifting kinetic energy.  For thawing fields, this means that the 
dynamics can actually turn around and head back toward the frozen, 
cosmological constant state.  
Recall that thawing fields start with $F_0=4/27$, so their maximum, 
turnaround value of $w$ occurs when $F$ has risen to $1/3$, at which point 
$w'=0$.  For freezing fields, the increase in 
$F$ simply moves them faster toward the cosmological constant. 

\begin{figure} [h!]
\includegraphics[width=  \columnwidth]{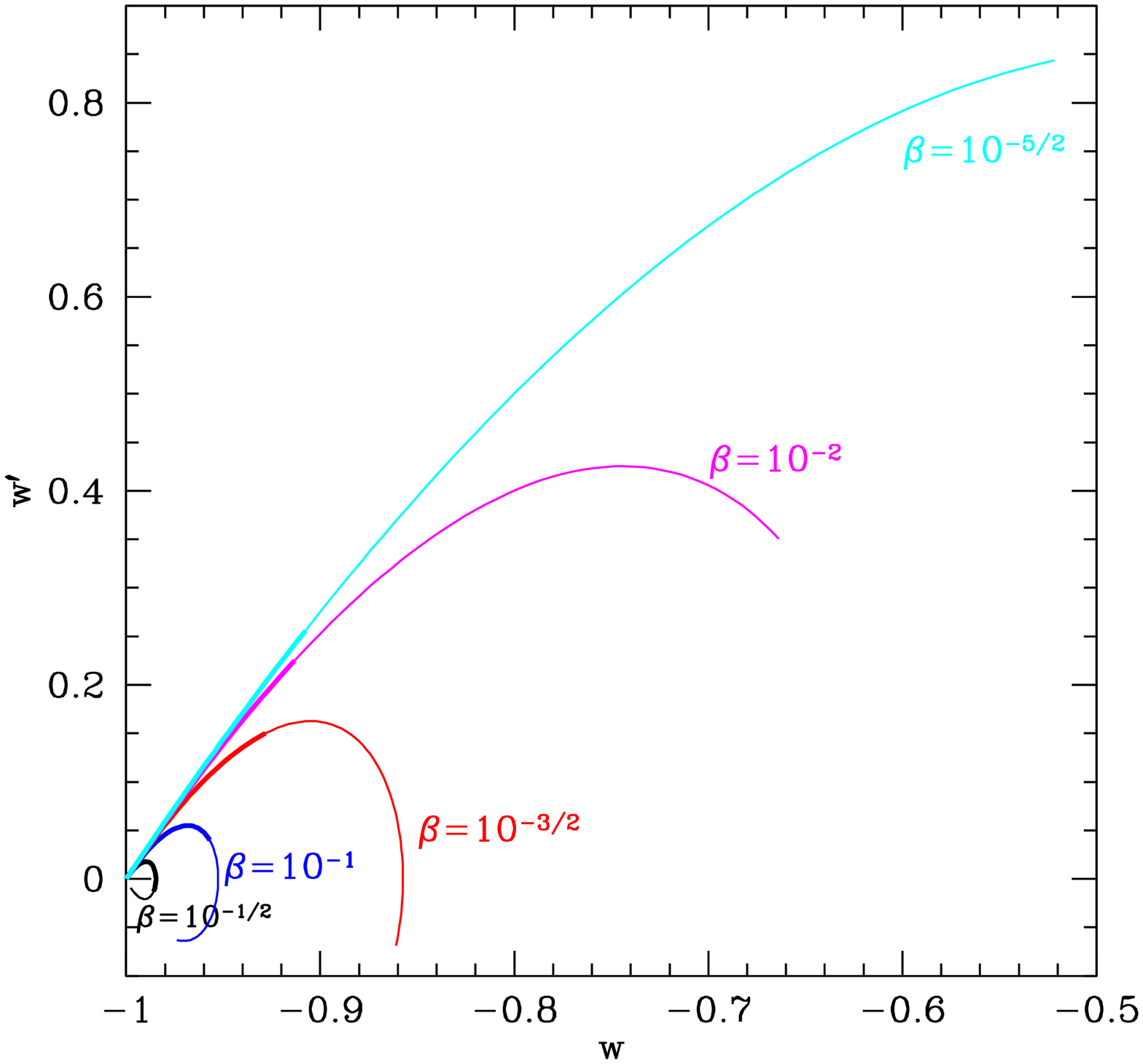}
\includegraphics[width=  \columnwidth]{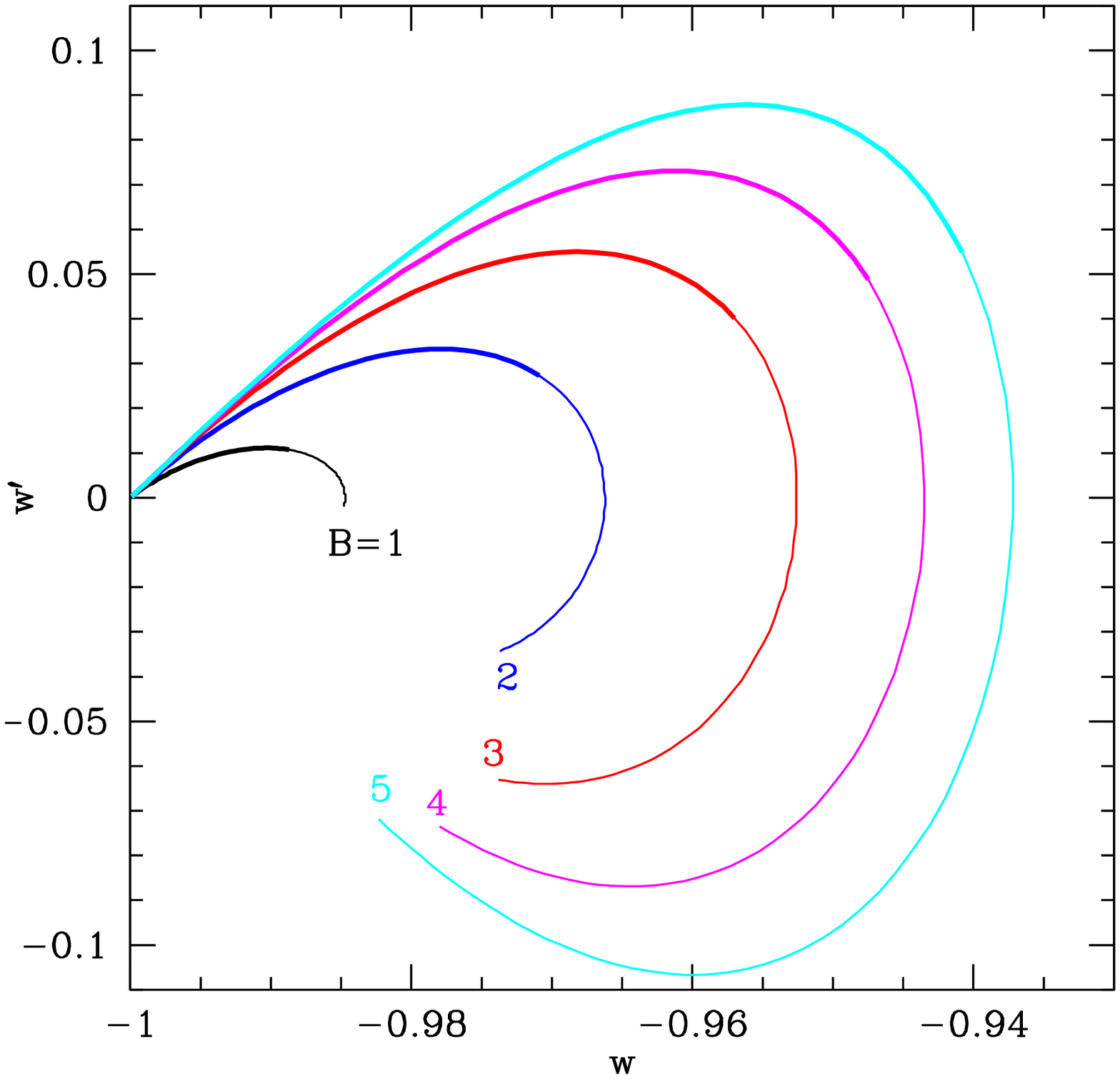}
\caption{[Top panel] The deviation amplitude of the flow parameter between 
the matter dominated era and today is given by $\beta$.  Increasing $\beta$ 
moves the dynamics away from small field deviations and toward skating, 
with rapid approach toward a frozen state.  [Bottom panel] For a fixed 
deviation amplitude, the transition rapidity is determined by $B$. 
Increasing $B$ postpones the transition from the matter dominated flow 
parameter, causing a more rapid evolution close to the present.  Both 
panels are for the thawing case, with $F_0=4/27$, and $1+w_i=10^{-4}$; 
curves extend to $a=2$ with thick portions of curves for $a\le1$.} 
\label{fig:varypart}
\end{figure}

\begin{figure} [h!]
\includegraphics[width= \columnwidth]{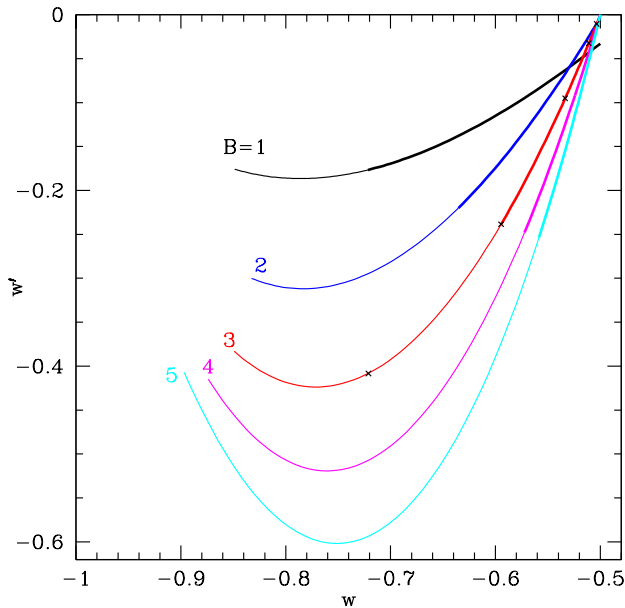}
\caption{As the bottom panel of Fig.~\ref{fig:varypart} but for the freezing 
case, with $F_0=1/3$, and $w_i=-0.5$.  Variation of $\beta$, as in 
Fig.~\ref{fig:varypart} top panel, tends to 
move the evolution along the curves; for example for $B=3$ we show by 
x's the $a=1$ location for $\beta=10^{-5/2}$, $10^{-2}$, $10^{-3/2}$, 
$10^{-1}$, $10^{-1/2}$, from right to left.} 
\label{fig:varyparf}
\end{figure}

Figure~\ref{fig:varypart}--\ref{fig:varyparf} show the effects of each 
parameter on the dynamics, for the thawing and freezing cases.  
The amplitude parameter $\beta$ controls the degree of evolution from 
the matter dominated state -- increasing $\beta$ has the same effect as 
increasing $F$.  The transition parameter $B$ governs the 
rapidity of evolution -- larger $B$ means that the transition from 
the matter dominated behavior happens nearer to the present.

\begin{figure} [h!]
\includegraphics[width= \columnwidth]{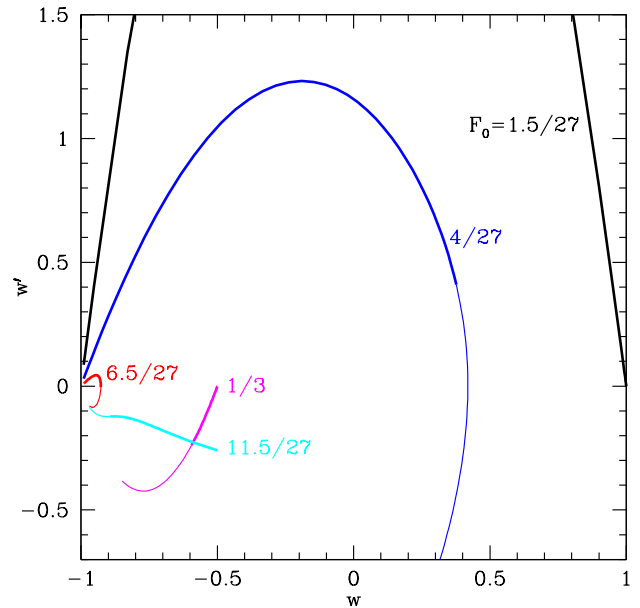}
\includegraphics[width= \columnwidth]{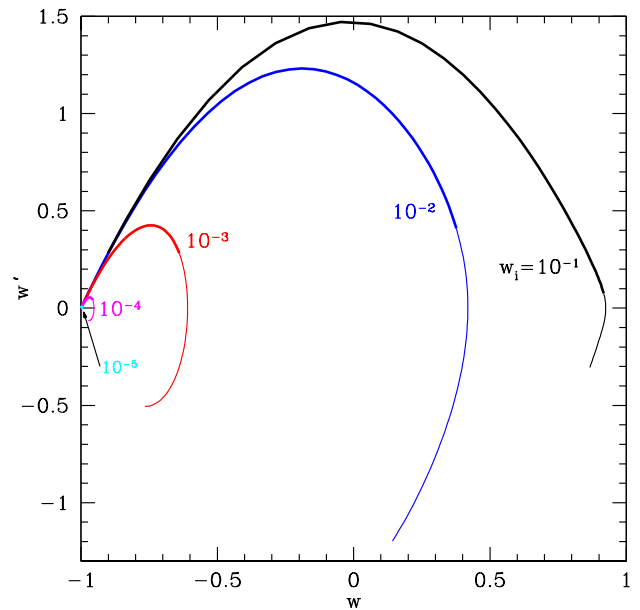}
\caption{[Top panel] The flow parameter during matter domination is 
generally determined to be $F_0=4/27$ or $1/3$, but other values are 
possible with fine tuning.  Non-standard values can cause unusual, 
strong dynamical behavior.  For $F_0\le 6.5/27$ we show curves taking 
$w_i=-0.99$; larger values of $F_0$ give unobservably small loops.  
For $F_0\ge 1/3$ we take $w_i=-0.5$.  
[Bottom panel] The dark energy equation of 
state at some fixed redshift in the matter dominated era can give a 
strong thrust to the evolution for thawing models if it is not near 
the thawing value $w_i=-1$ expected from Hubble drag.} 
\label{fig:f0wi}
\end{figure}

The initial condition parameters $F_0$ and $w_i$ have a significant 
influence, as seen in Fig.~\ref{fig:f0wi}.  Increasing $F_0$ gives 
the dynamics a significant push in its evolution.  If $F_0\ne4/27$ 
then the field somehow ignores the high Hubble friction during matter 
domination.  Basically the field must have been ``super-accelerating'', 
i.e.\ $\ddot\phi\gg H\dot\phi$, if $F_0<4/27$, while it must have been 
fine tuned to be coasting $\ddot\phi\ll H\dot\phi$ (not slow-roll since 
the matter, not the dark energy, dominates the expansion) if 
$4/27<F_0<1/3$ (cf.~\cite{paths}).  The value $F_0=1/3$ corresponds to 
the dynamics determined by an attractor solution for the field, but 
$F_0>1/3$ indicates ``super-deceleration'', $-\ddot\phi\gg dV/d\phi$, 
equivalent to fine tuning for a ``super-flat'' potential, yet somehow 
the field still rolls.  These properties motivate why we will later 
consider only the values $F_0=4/27$ and $1/3$.

Similarly, increasing $w_i$ for thawing fields gives a strong boost 
to the evolution, pushing the dynamics far from $w=-1$.  (For tracker 
freezer fields, $w_i$ is determined by the attractor solution and 
then $w'_i=0$.)  
For thawing fields, we expect $1+w_i$ to be of order $\Omega_{\phi,i}$, 
the dark energy density at the initial redshift considered in the matter 
dominated era (see, for example, \cite{cahndl}).  (Taking $a_i=0.1$, 
typically $w_i\approx-1+10^{-4}$.)  For freezing fields $w_i$ 
can be of order unity, but again tracking evolution starts with $w'=0$.  
Another aspect of non-standard values of $F_0$ or $w_i$ is the presence 
of excessive early dark energy density.  For example, a thawing field 
with $1+w_i=10^{-2}$ has $\Omega_{\phi,i}\approx0.06$, which would cause 
conflict with growth of large scale structure and the cosmic microwave 
background observations \cite{edelimit}.  A thawing field 
with $F_0=1.5/27$ would break matter domination completely. 

Thus, we effectively have a three parameter description, given by 
the set $\{\beta,B,w_i\}$, for each of two cases: $F_0=4/27$, $1/3$. 
In general even a three parameter description of dark energy dynamics 
cannot be fully constrained even with next generation data 
\cite{linhut05,deplpca}.  
Indeed, determining the transition time or speed for the dark energy 
equation of state is extremely difficult \cite{linhut05,rapetti,dynq} 
so we do not expect $B$ to be constrained.  However, this form is useful 
because it gives considerable freedom for dynamics while at the same 
time  taking into account the oldness of dark energy, i.e.\ the long 
influence of matter domination on the evolution. 

The ansatz also provides for three classes of dynamical behavior: the 
usual thawing and freezing classes, and ``re-freezing'' models where 
the field thaws from the Hubble friction-induced torpor, evolves to a 
maximum $w$, but then can enter the freezing region and eventually evolve 
to a potential minimum and cosmological constant state.  This last class 
will be especially interesting as it bounds the area of the $w$-$w'$ 
plane that lies within a certain observational distance of $\Lam$, i.e.\ 
models that agree with the $\Lam$CDM expansion history to within a 
certain precision (see the next section).  One can imagine a fourth 
class which starts as 
trackers but then turns away from the cosmological constant, i.e.\ 
somehow moves away from the minimum, but there is little physical motivation 
for this and we do not consider it further.

\section{Constraining Dynamics \label{sec:condyn}} 

In the previous section we saw that rich dynamics is still available 
to the dark energy field, even after taking into account its oldness -- 
the effect of the long era of matter domination to age and mellow the 
arbitrariness of the early evolution.  In particular, we emphasized 
the naturalness of the age influence and did not impose any restrictions 
on behavior such as assuming small field excursions all the way until 
late times.  Here we consider how observational data will be able to 
constrain the later time behavior within the flow parameter formulation 
and narrow in on regions of the $w$-$w'$ plane. 

\subsection{Scanning Phase Space \label{sec:phase}} 

In fact, we will find clear, bounded regions of the phase space and 
maximum allowed deviations in $1+w$, assuming some future 
distance data.  To scan over all the possible dynamics within the 
flow ansatz we begin by considering the parameter ranges.  As discussed 
in the previous section, there are good reasons from both theory and 
observations to consider {\it initially\/} thawing and tracker/freezer 
classes, with $F_0=4/27$ and $1/3$ respectively.  We emphasize that 
these values only give the initial conditions deep in the matter 
dominated era and are not assuming later behavior.  

For $1+w_i$, we indicated that for thawing models this was often around 
$10^{-4}$, while for freezing models it could lie between a small value 
and 1, though it is more difficult to achieve $w\approx-1$ today if 
$w_i$ is not sufficiently negative.  However, to keep open all 
possibilities, we consider $1+w_i$ ranging between $10^{-5}$ and 1, 
with a log prior for the thawing class. 
For the amplitude $\beta$, most models generally considered have values 
between few$\times10^{-2}$ and few$\times10^{-1}$.  Again to be flexible, 
we take a log prior between $10^{-5/2}$ and 1.  The evolution rapidity 
$B$ can be examined from the first order correction (in the small dark 
energy density) to matter domination; for thawing models $B=3$ and for 
tracking models $B=-3w_i$.  Numerical solutions show that for many thawers 
$B$ slowly declines toward the present, reaching $\sim1.5$, and for 
trackers $B$ slowly rises toward the present, reaching $\sim2$.  We 
take a generous linear prior on $B$ between 1 and 5. 

The models obtained by scanning over the parameter space on a 
$50\times50\times50$ grid in $\beta$-$B$-$w_i$, for $F_0=4/27$ and 1/3,  
are then compared in the distance-redshift relation to the distances to 
redshifts $z=0.1$, 0.2,\dots 1.7 taking a 
$\Lambda$CDM fiducial cosmology.  
The 250,000 generated models are 
referred to as the prior sample, and those that pass observational 
cuts are referred to as the viable sample.  The observational cut 
considered first is simply requiring the distances at each redshift 
to lie within 1\% of fiducial (roughly next generation data precision), 
and the dark energy density at $a_i=0.1$ 
must obey $\Omega_{\phi,i}<0.03$.  We later consider a more sophisticated 
Monte Carlo likelihood calculation. 

\begin{figure} [h!]
\includegraphics[width= \columnwidth]{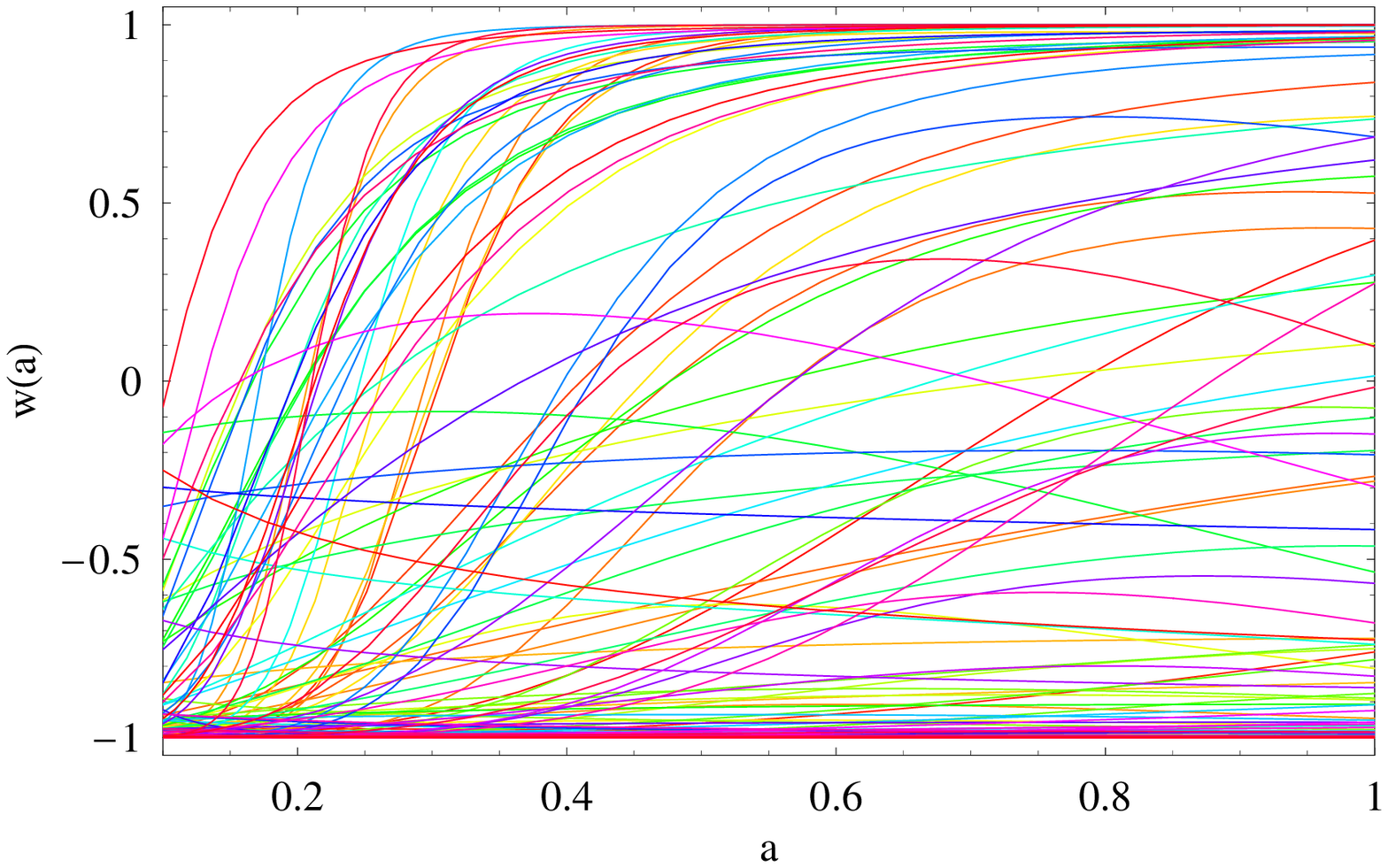}
\includegraphics[width= \columnwidth]{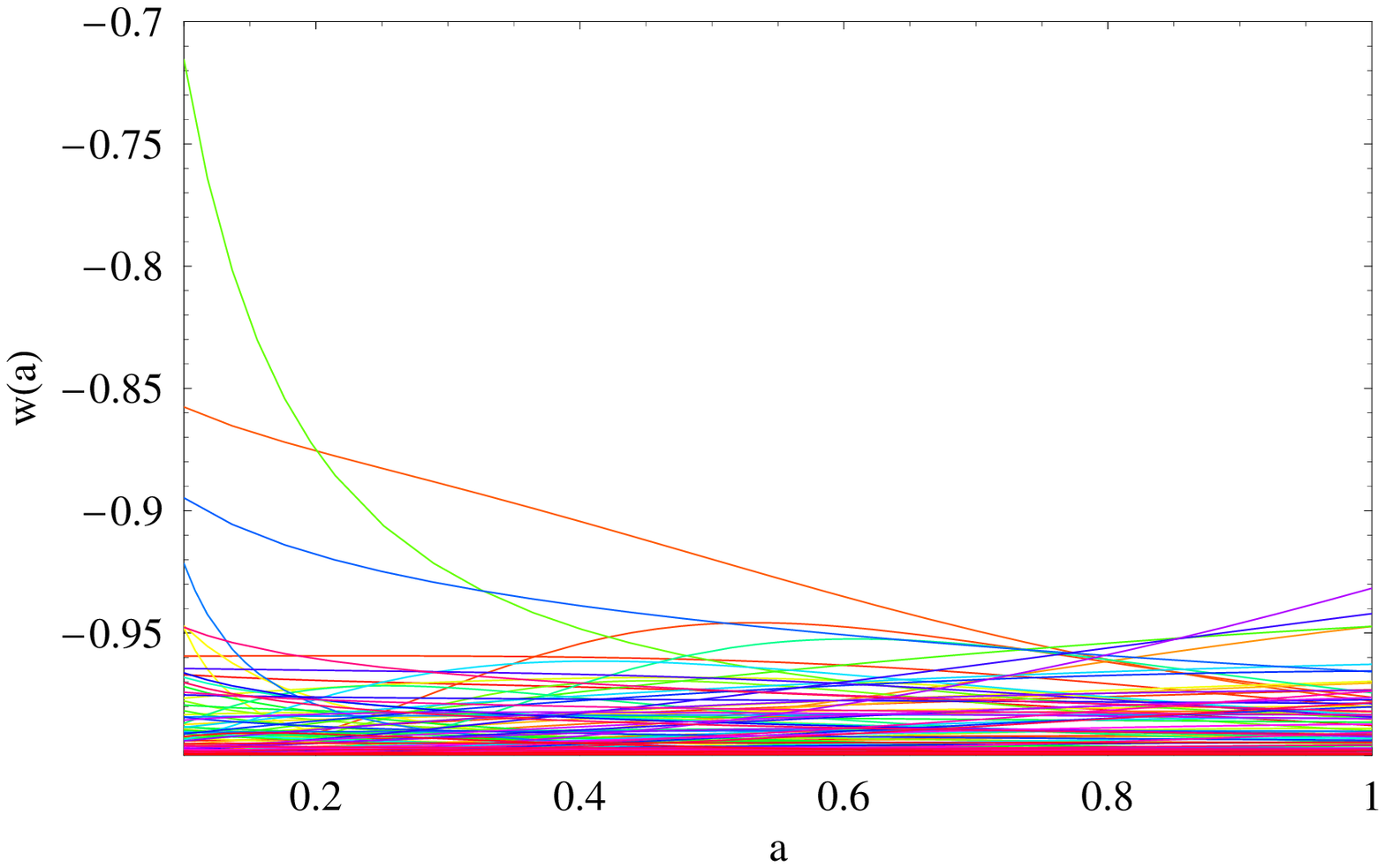}
\caption{The flow ansatz allows a rich variety of evolutionary behavior, 
including nonmonotonicity, while incorporating the physical influence 
of the long matter dominated era.  The top panel shows a random selection 
of 150 models (note many are very close to $w=-1$) while the bottom panel 
shows 150 random models that pass the observational cuts.  (Note the 
different vertical scales.) Thawing, freezing, and nonmonotonic evolution are 
all represented.} 
\label{fig:prepost} 
\end{figure}

Figure~\ref{fig:prepost} shows the dynamics in terms of the evolution of 
the equation of state with scale factor, $w(a)$, 
for the prior and viable models.  While the observations certainly 
constrain the viable models to more limited {\it values\/} of the equation 
of state, the variety of dynamics allowed by the flow parametrization is 
still diverse.  In particular, models with thawing (leaving $w=-1$), 
freezing (approaching $-1$), and refreezing (nonmonotonic in $w$) behavior 
are all represented. 

\begin{figure} [h!]
\includegraphics[width=  \columnwidth]{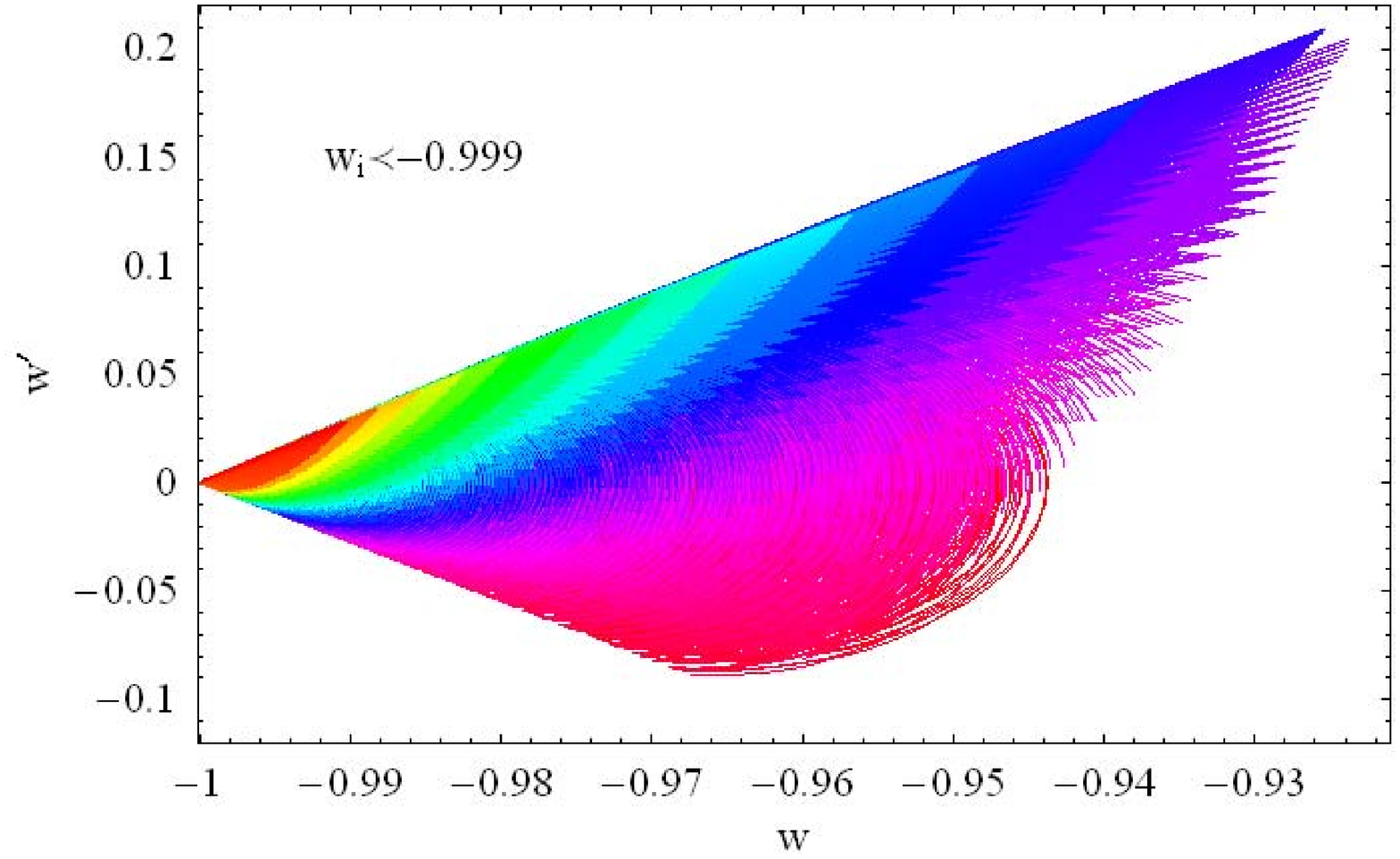} 
\includegraphics[width=  \columnwidth]{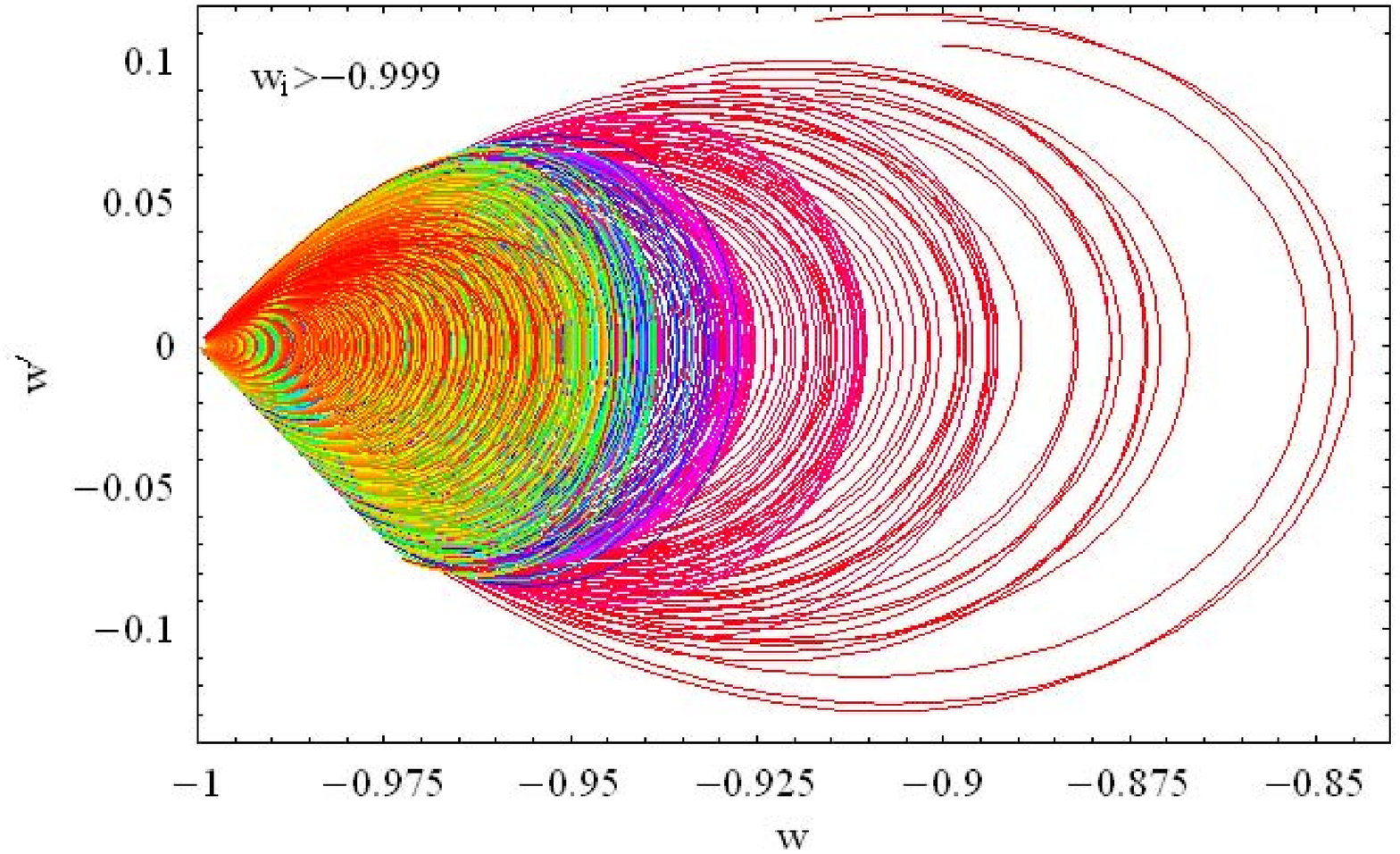}
\caption{Models deviating from $\Lambda$CDM distances by less than 
1\% live in a well defined region of phase space, here shown for 
the initially thawing class.  Colors/shading 
represent different values of $w_i$, with models possessing the 
smallest values of $1+w_i$ restricted to the bright, red region 
closest to $w=-1$.  The top panel shows models with $w_i<-0.999$ 
(as expected for most thawing models); the bottom panel shows 
models with $w_i>-0.999$.  The sparseness of the bottom panel is 
due to the finite gridding: only a tiny percentage of the 2500 models 
for each $w_i$ pass the distance criteria.  The maximum equation of 
state deviation allowed at any redshift corresponds to 
$w_{\rm max}=-0.845$.} 
\label{fig:wwp01pctt} 
\end{figure}

\begin{figure} [h!]
\includegraphics[width=  \columnwidth]{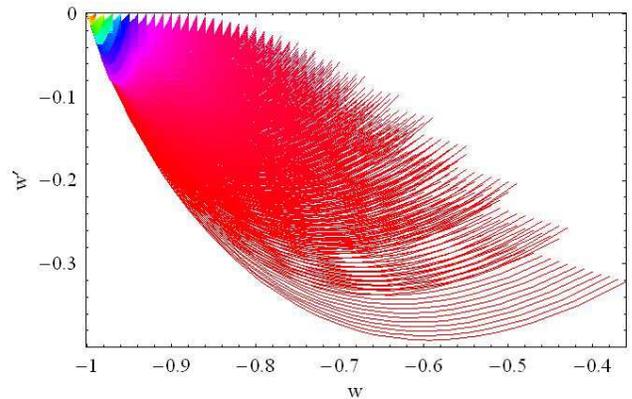}
\caption{As Fig.~\ref{fig:wwp01pctt}, but for the initially freezing 
class.  The  maximum initial equation of state deviation, at $a_i=0.1$, 
corresponds to $w_{\rm max}=-0.36$, reaching $w_0=-0.985$ today.} 
\label{fig:wwp01pctf}
\end{figure}

Observational constraints translate to well defined regions in the $w$-$w'$ 
phase space, with a maximum allowed distance deviation imposing a 
maximum allowed equation of state deviation $w_{\rm max}$.  Furthermore, 
the allowed region is compact, limiting $w'$ as well.  
Figures~\ref{fig:wwp01pctt}-\ref{fig:wwp01pctf} shows the regions 
that have less than 1\% 
distance deviation from the $\Lambda$CDM fiducial, with the colors/shading 
representing different values of $w_i$, as well as scanning over $\beta$ 
and $B$.  Note that in Fig.~\ref{fig:wwp01pctt} for small $1+w_i$ most 
models look like regular thawing 
models, while as $1+w_i$ increases, the only way the models can obey 
the observational constraints are to loop around as refreezing models.

\subsection{Mirage Model Dynamics \label{sec:mirage}} 

\begin{figure}[h!]
\includegraphics[width=  \columnwidth]{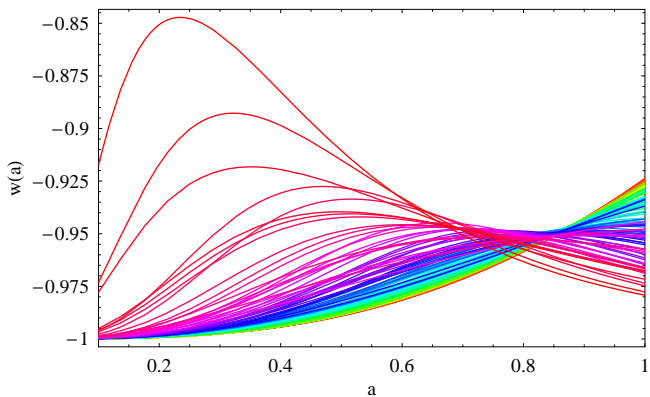}
\caption{Initially thawing models that have maximum distance deviations 
from $\Lambda$CDM by some fixed amount, here 0.999--1\%, have a variety 
of behaviors but exhibit a narrow waist at $a\approx0.8$.  In terms of 
a constant or averaged equation of state, these will all look like 
$\langle w\rangle\approx-0.95$ but have distinct physics. } 
\label{fig:thaw1pct}
\end{figure}

\begin{figure} [h!]
\includegraphics[width= \columnwidth]{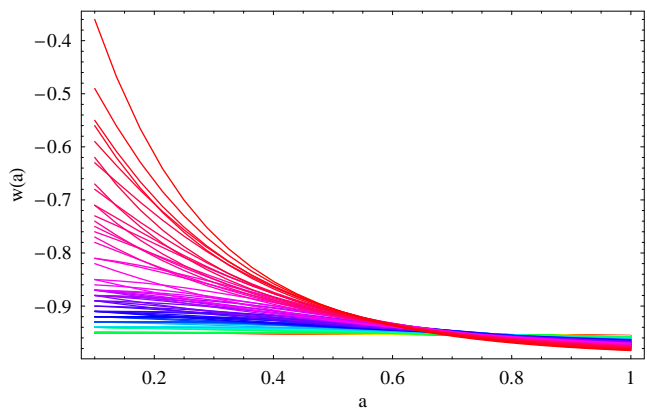} 
\caption{As Fig.~\ref{fig:thaw1pct} but for freezing models. Note the 
waist persists, though here it is at $a\approx0.7$. }
\label{fig:frz1pct}
\end{figure}

An interesting special case to consider is 
those models that deviate from $\Lambda$CDM by exactly 
1\%, rather than those in the whole range 0-1\%.  These could represent, 
for example, a specific fiducial distinguishable from $\Lambda$CDM 
by next generation data.  Figures~\ref{fig:thaw1pct}-\ref{fig:frz1pct} 
show the $w(a)$ evolutionary behaviors.  Of particular note is the waist 
near $a=0.8$, where the equation of state has a dispersion of only 
$\Delta w\approx0.01$.

This sort of waist feature has been discussed before 
in the context of mirage models \cite{mirage} and pivot points, albeit 
restricted to the parametrization $w(a)=w_0+w_a(1-a)$.  Such 
a characteristic highlights the importance of experiments capable of 
measuring the time variation $w'$ of the equation of state; a survey 
only able to measure a coarse time average will ineluctably find 
$\langle w\rangle\approx0.95$ for all these diverse models.  If the 
dynamical model allowed the equation of state to become more negative than 
$-1$, then one could easily obtain a waist with $\sim$0\% distance 
deviation from $\Lambda$CDM and find $\langle w\rangle=-1$, wrongly deducing 
a cosmological constant as the answer.  This is the mirage aspect of 
the mirage models, and so uncovering the true physics behind dark 
energy requires an experiment with a long redshift baseline over which 
to measure the time variation of the equation of state.

\subsection{Dynamics Today \label{sec:wtoday}} 

\begin{figure} [h!]
\includegraphics[width= \columnwidth]{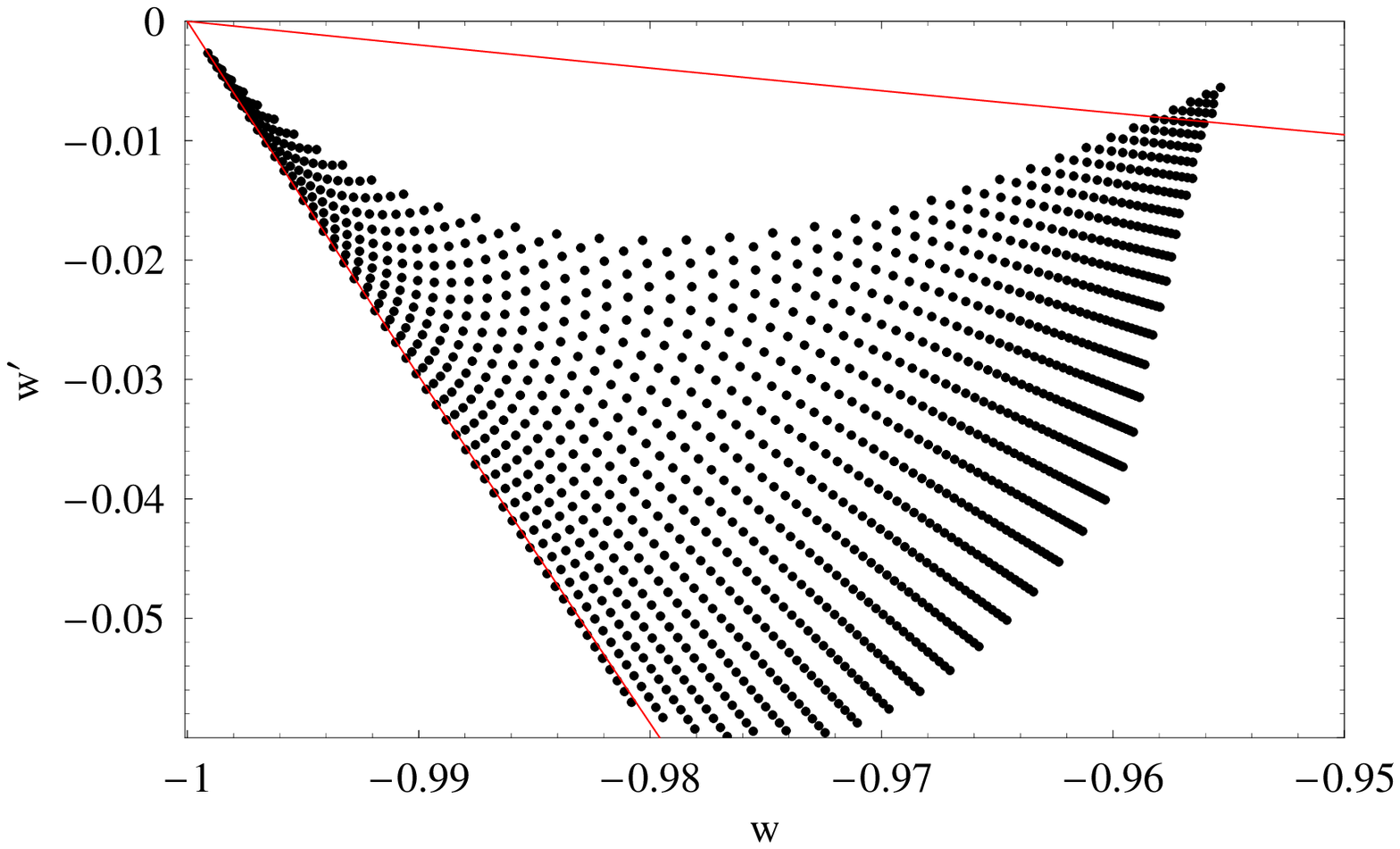}
\includegraphics[width= \columnwidth]{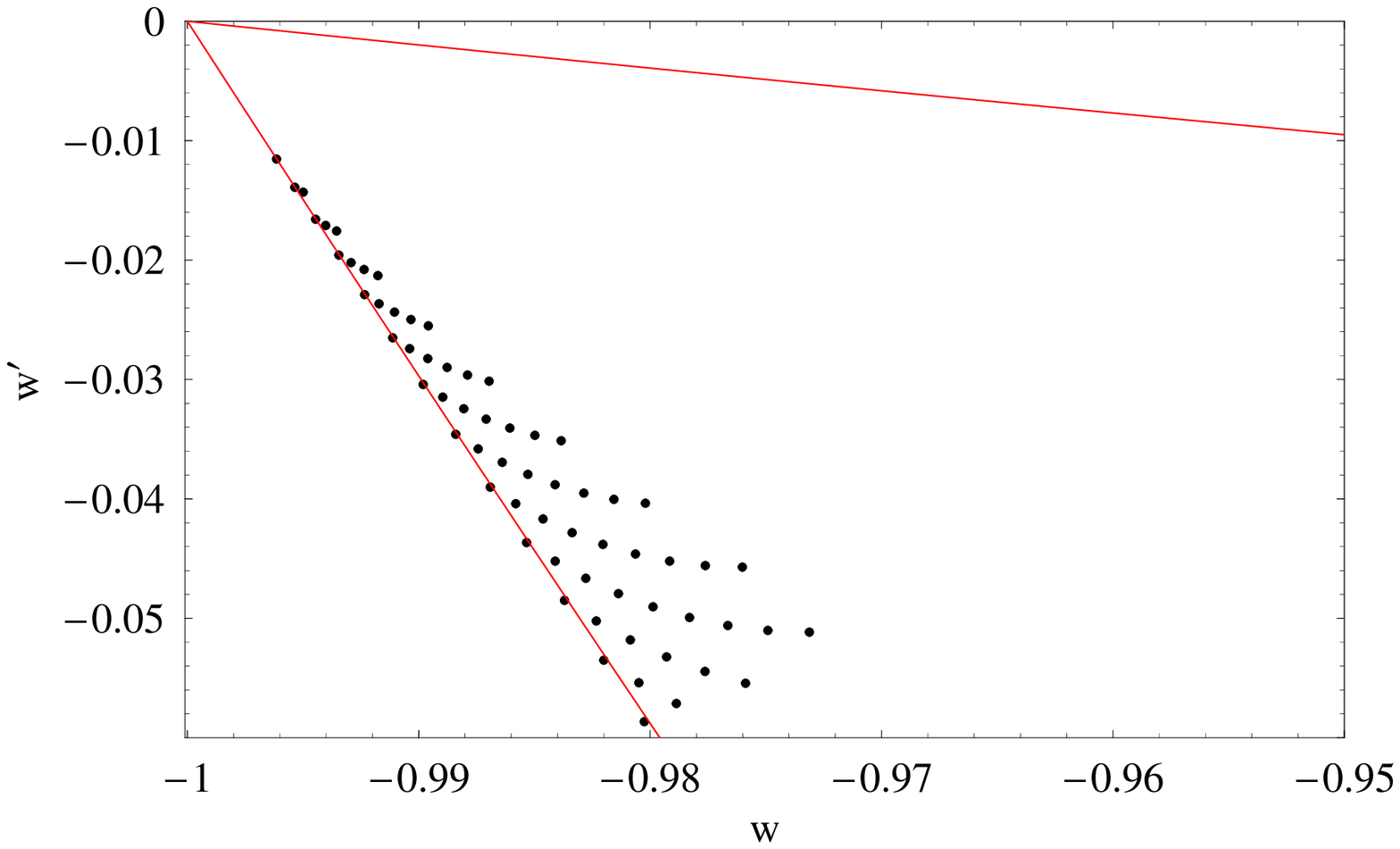}
\caption{The dynamics today, $w_0$-$w'_0$, are shown with each dot 
representing 
the present endpoint of a trajectory in Fig.~\ref{fig:wwp01pctf} for 
freezing models giving distance deviations of 0--1\% from $\Lambda$CDM. 
As $w_i$ moves away from $-1$, the region gradually grows, reaching 
a maximum extent at $w_i\approx-0.95$ (top panel), then shrinking and 
coalescing along the bottom boundary of the freezing region (delimited 
by the straight lines).  The bottom panel shows the case of $w_i=-0.8$. } 
\label{fig:wwp0f}
\end{figure}

We can also consider the properties of the dark energy today. 
For initially thawing models, the values of $w_0$ and $w'_0$ can 
lie basically anywhere within the overall observationally allowed 
region, e.g.\ the top panel of Fig.~\ref{fig:wwp01pctt} (for the 
bottom panel, where $w_i>-0.999$, the values today lie at the end 
of the loops, within the top panel region).  For initially freezing 
models, the distribution today is more interesting.  For $w_i$ 
near $-1$, the values today fill most of the appropriate part of the 
freezing region in Fig.~\ref{fig:wwp01pctf}.  The region allowed 
by 1\% distance observations is largest when $w_i\approx-0.95$. 
As $w_i$ increases, however, the dynamics today hugs the lower 
boundary of the freezing region, $w'=3w(1+w)$, and the allowed 
region shrinks, vanishing for $w_i>-0.36$. 
Figure~\ref{fig:wwp0f} illustrates these characteristics.

\begin{figure} [!h]
\includegraphics[width=  \columnwidth]{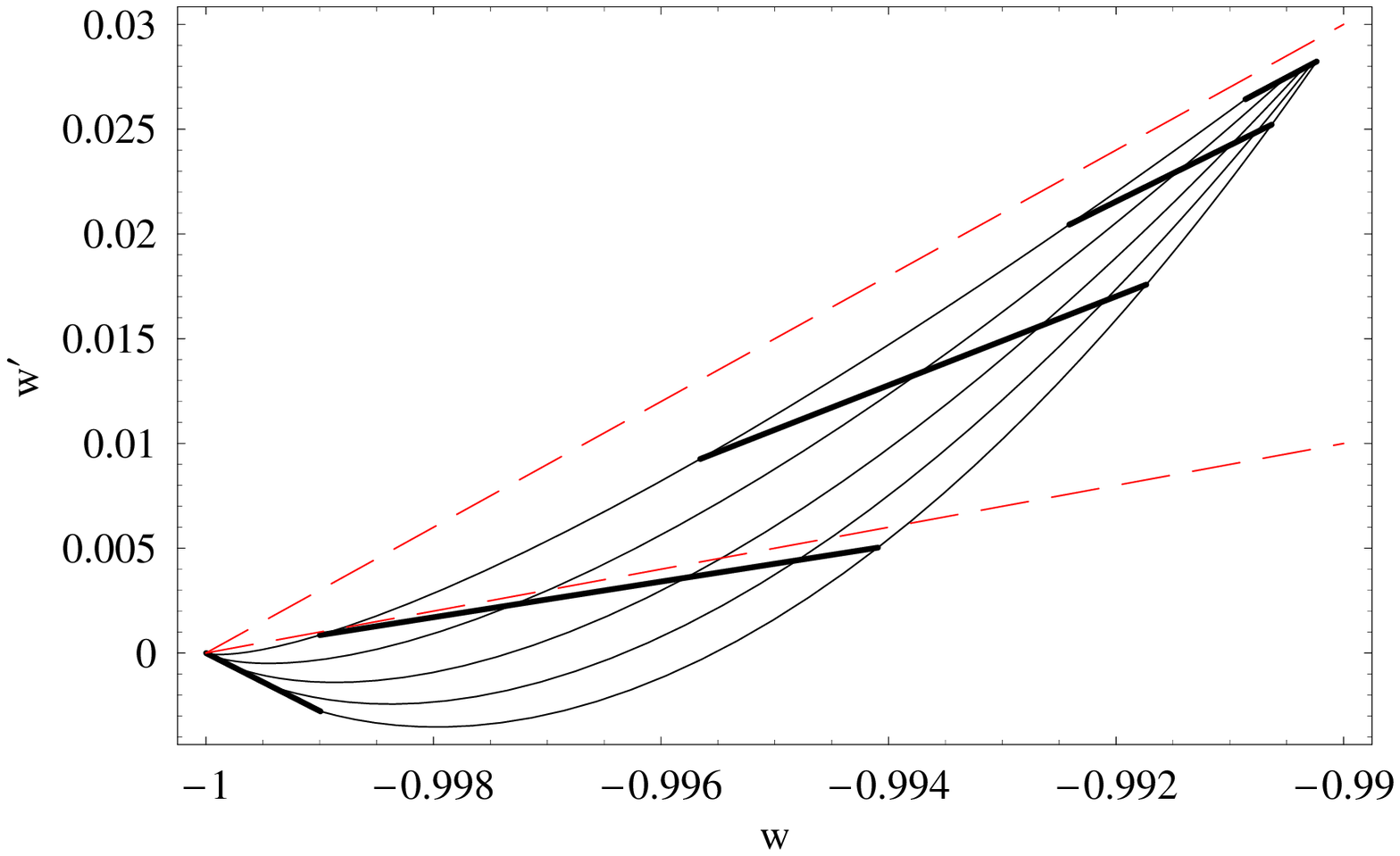}
\includegraphics[width=  \columnwidth]{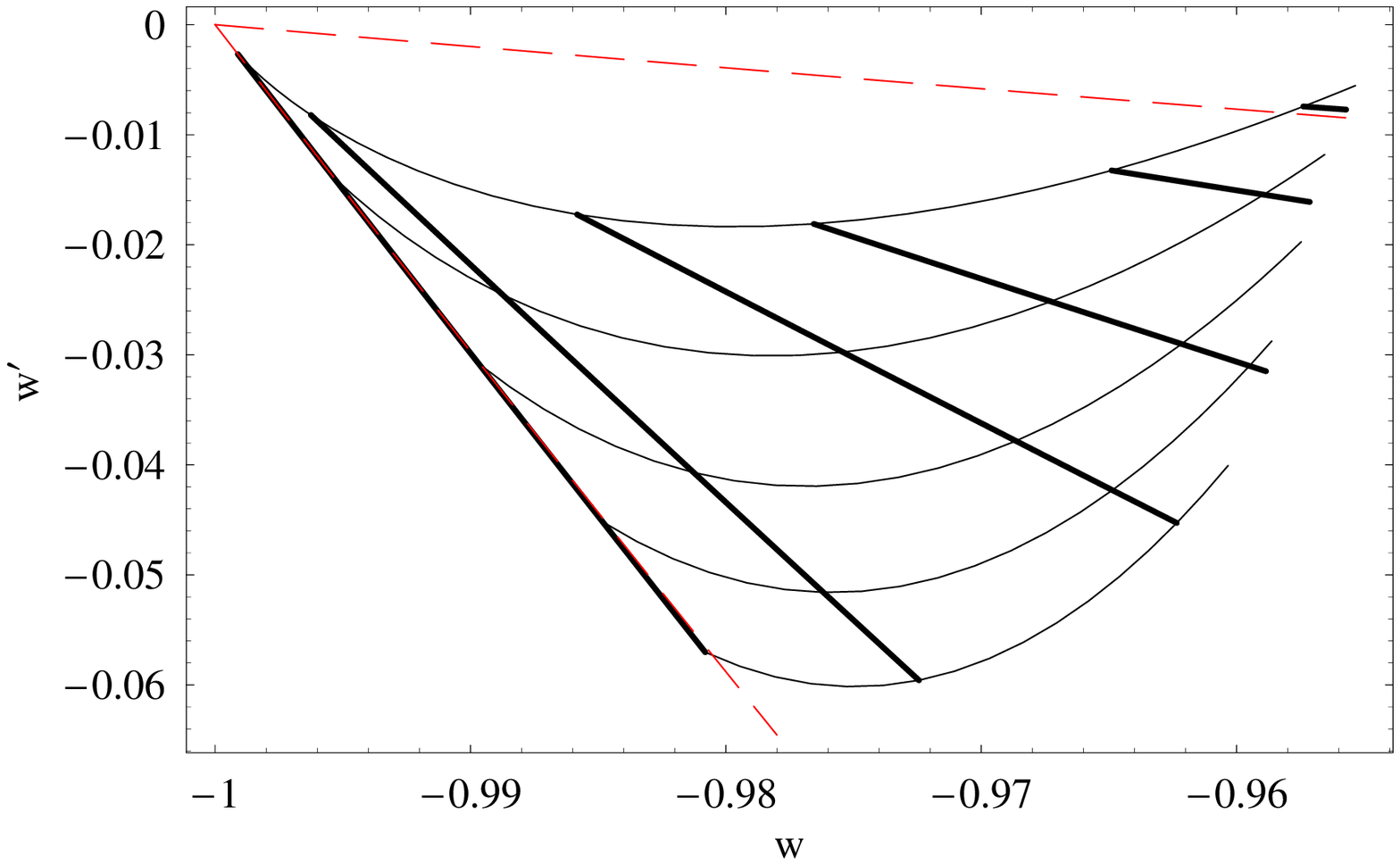}
\caption{For the present values of the equation of state parameters 
$w_0$ and $w'_0$, the influence of flow parameters $\beta$ and $B$ 
are illustrated.  The isocontours of $\beta$ (thick lines) are shown 
for $\beta=1$, 0.11, 0.033, 0.01, 0.0032 from bottom to top for the 
top panel ($\beta=1$, 0.5, 0.2, 0.11, 0.05, 0.02 from left to right 
for the bottom panel).  Isocontours of $B$ (thin lines) are 
shown for $B=1$, 2, 3, 4, 5 from top to bottom.  
The top panel shows the initially thawing case, the bottom panel 
the initially freezing case, with the dashed lines showing the 
conventional boundaries for each region. } 
\label{fig:iso} 
\end{figure}

To understand the elements of the dynamics, we show in Fig.~\ref{fig:iso} 
the role of the parameters $\beta$ and $B$ on the present value of 
the equation of state and its time variation, holding $w_i$ fixed.  
As expected, $\beta$ is more influential than $B$, with large values 
of $\beta$ accelerating the evolution toward a present state near 
the cosmological constant.

The original classification of the boundaries of the thawing and 
freezing regions \cite{caldlin} was made by studying numerical 
solutions for a range of field potentials and determining their 
endpoints, i.e.\ $w_0$-$w'_0$ values today.  Although the long 
period of matter domination guides the flow approach dark energy 
models toward an initial behavior either along the upper boundary 
of the thawing region, $w'=3(1+w)$, or to the tracker condition of 
$w'=0$ and subsequent approach toward the cosmological constant, 
we can now address quantitatively whether their evolution 
leads to consistency with the thawing and freezing regions today. 

We clearly see from Fig.~\ref{fig:wwp01pctf} that the initially 
freezing models lie through most of their evolution in the canonical 
freezing region $0.2w(1+w)\le w'\le 3w(1+w)$, and from Fig.~\ref{fig:wwp0f} 
that in particular they all lie in the freezing region today.  
Recall though that thawing models in the flow approach can either 
remain thawing, loop around, or even refreeze.   Out of all the viable 
(0-1\% distance deviation) initially thawing models, 47\% remain in 
the thawing region through the present, while 40\% today lie in what 
would usually be the freezing region, and 13\% today lie 
between the thawing and region regions near the coasting 
($\ddot\phi=0$) line.  If we restrict consideration to the most 
natural initially thawing models, those with $1+w_i<5\times10^{-4}$ 
(recall most standard thawers have $1+w_i\approx10^{-4}$), then the 
percentages become 60\%, 28\%, 12\% respectively. 
Thus there is no shortage of thawing models today.

\subsection{Monte Carlo Constraints \label{sec:mcmc}}

In the previous subsections we made use of a grid tessellation of 
parameter space to examine the variety of $w$-$w'$ phase space 
behaviors that can be obtained within the flow formalism.  Here we 
will probe constraints on the $F(a)$ parameter space.  In particular, 
we are interested in the ability to distinguish between model classes 
given observational data, and secondarily to determine the values of the 
flow parameters.  
We focus on the capability of future 
observations to rule out the cosmological constant, for example 
through measuring an evolution through a nonzero $\beta$ or finding 
the high redshift ($z\gtrsim 10$) equation of state value $w_i$ different 
from $-1$. 

To quantify this, we generate data using a fiducial model and perform 
a Markov Chain Monte Carlo (MCMC) likelihood analysis 
\cite{mackay,gilks,cosmomc,dunkley_corr,liddle_mcmc} on the $F(a)$ 
parameters which, together with the initial value of the equation of 
state parameter $w_i$, fully determine the evolution of the dark energy. 
The usual $\Lam$CDM model corresponds to $w_i=-1$ and the best fit is 
achieved with a set of parameters that select from the prior the 
smallest value of $1+w_i$ (to start close to $\Lam$), the largest of 
$\beta$ (to make the field quickly freeze, i.e.\ approach $\Lam$), and 
the smallest value of $B$ (to start the freezing evolution early).  
Basically the cosmological constant behavior is an asymptotic state for 
scalar field dynamics.  Because of this, the likelihood would concentrate 
around the prior bounds for whatever ranges we choose, making the 
predictions prior dependent.  To avoid this, we instead choose a 
fiducial cosmology distinct from $\Lam$CDM. This is of further interest 
since it explores the sensitivity of $F(a)$ parameters in revealing this 
deviation, should next generation datasets narrow in on a region of 
parameter space off $\Lam$CDM. 

Following the approach of the previous sections we study thawing and 
freezing models separately since these have well defined, high-redshift 
responses to the flow function $F(a)$.  Accordingly, we select $F_0$ 
equal to $4/27$ and $1/3$, respectively.  As discussed in 
Sec.~\ref{sec:fform}, fields with initial $F$ values different from 
these either betray matter domination, dilute very fast, or are finely 
tuned coasters.  Nonetheless we emphasize that $F_0$ merely determines 
the high redshift initial conditions of the field and other parameters 
govern its late time dynamics.  The distinction between thawing and 
freezing origin motivates further a different choice of priors for 
$1+w_i$, since the different initial behavior of $w(a)$ is one of the 
features distinguishing between the two classes of models. For thawers, 
$1+w_i$ is related to the initial energy density $\Omega_{\phi,i}$ and 
can be quite small, so we choose a logarithmic prior on $1+w_i$ for the 
thawing models; for trackers, $1+w_i$ tends to be larger so we choose a 
uniform prior on $1+w_i$ for the (tracker) freezer case.  
Similarly, since $\beta$ tends to be larger in the freezing case we 
choose a uniform prior there but a logarithmic prior for the thawer 
case.  The bounds on the priors, however, are the same as for 
the previous grid scanning: $1+w_i\in[10^{-5},1]$, 
$\beta\in[10^{-5/2},1]$, and $B\in[1,5]$.  

The fiducial cosmology for both classes of models is chosen to be a 
maximum distance deviation of $\sim 1\%$ away from $\Lam$CDM, to test 
the constraint capability of next generation observations.  We assume 
data of 1\% accuracy in distance at redshifts $z=0.1$, 0.2,\dots 1.7.  
This allows us to explore the degree to which the data can distinguish 
between the fiducial (thawing or freezing), its opposite class (freezing 
or thawing), and the cosmological constant.  When considering a thawing 
fiducial we adopt $F_0=4/27$, $1+w_i=10^{-4}$, $\beta=0.04$, and $B=2$.  
For a freezing fiducial we take $F_0=1/3$, $w_i=-0.75$, $\beta=0.6$, 
and $B=1.5$.  

The MCMC sampling of parameter space is done with the Metropolis-Hastings 
algorithm, using a Gaussian proposal distribution and fours chains per case. 
Computation of 
probability distributions and likelihood contours is done using a 
modified version of the GetDist package, provided with the CosmoMC 
software \cite{cosmomc}.  We define $w_i=w(a_i)$ with $a_i=0.1$ and 
define the present by $\Omega_{\phi,0}=0.72$. 
We integrate the coupled equations 
\beqa 
w'&=&-3\,(1-w)^2\,\left[1-\frac{1}{\sqrt{3F}}\right] \\ 
\ophi'&=&-3 w \ophi\, (1-\ophi) \\ 
d'&=&a^{-1}\,\sqrt{\frac{1-\ophi}{(1-\Omega_{\phi,0})\,a^{-3}}}\,, 
\eeqa 
where $d(a)$ is the conformal distance and a prime denotes $d/d\ln a$.  
The likelihood is based on 
the comparison of $d(a)$ for a $F(a)$ model to $d(a)$ for the fiducial.

\begin{figure*} [t!]
\includegraphics[width= \columnwidth]{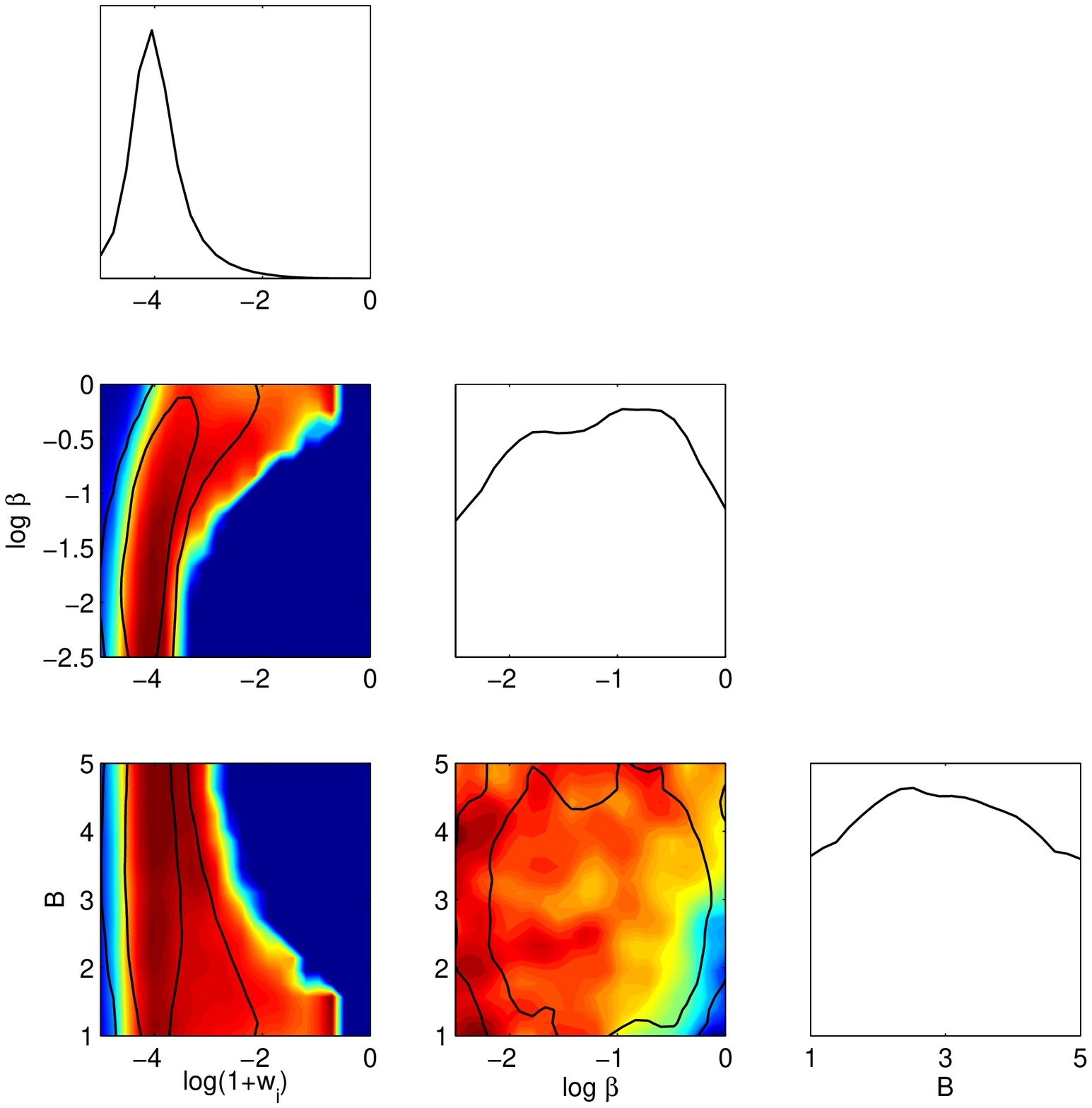}
\includegraphics[width= \columnwidth]{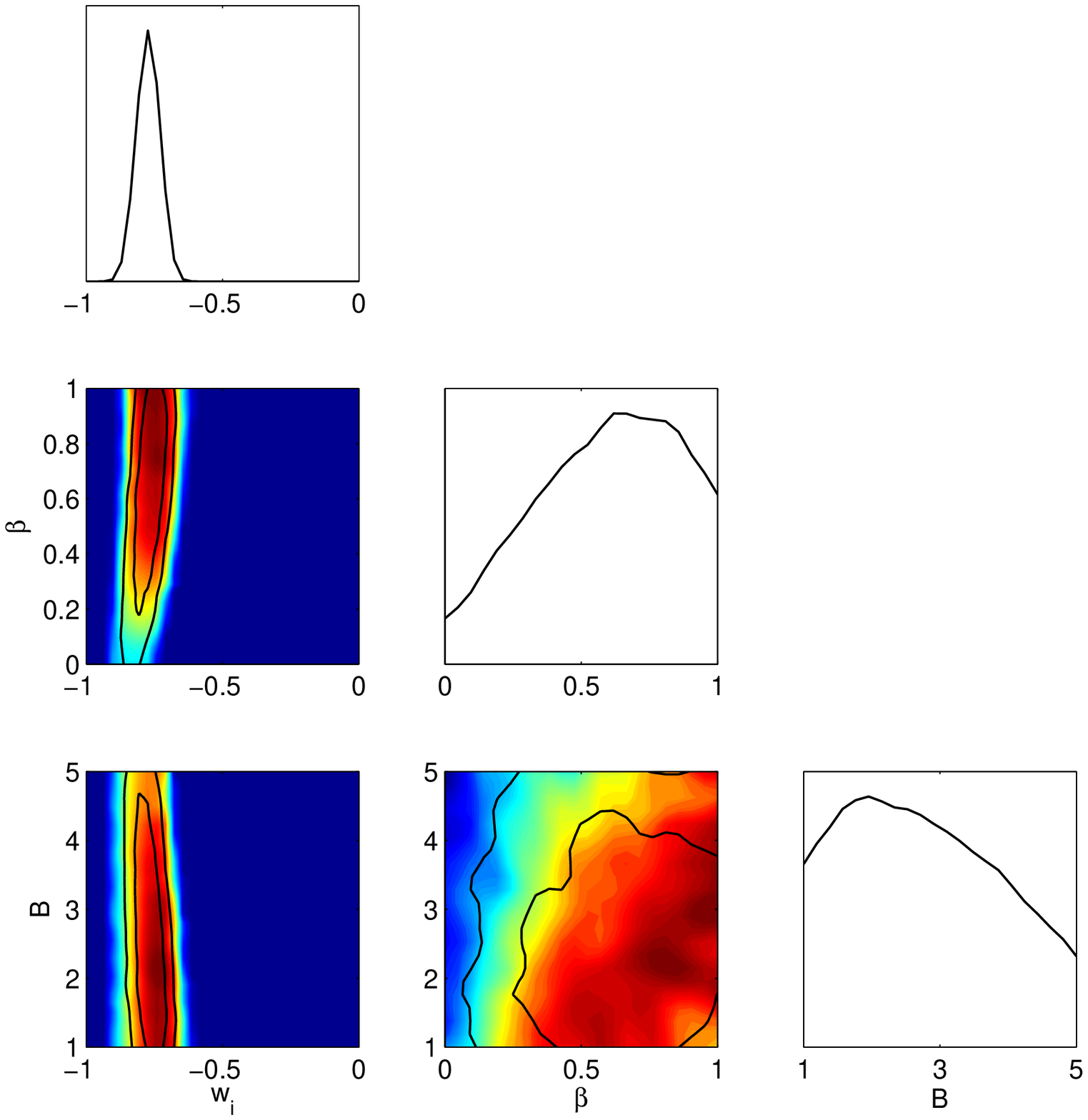} 
\caption {1- and 2-dimensional marginalized probability distributions 
and contours of MCMC reconstruction of the flow parameters are 
illustrated for next generation distance data. 
The left panel shows results for the thawing case, with $F_0=4/27$, 
which used log priors on $1+w_i$ and $\beta$; the right panel shows the 
distributions for the tracker/freezing case, with $F_0=1/3$, which 
used uniform priors to take into account the larger fiducial values. 
The diagonal panels correspond to the marginalized 1D probability 
distributions and the off diagonal to the marginalized 2D joint 
distributions. 
Thick curves in the off diagonal plots show the $68\%$ and $95\%$ likelihood 
contours and the color grading denotes the marginalized probability 
density from (red) highest to (blue) lowest.  
The MCMC constraints clearly recognize small $1+w_i$ for thawing, 
distinct from the large values typical of freezing models and from the 
zero value of the cosmological constant.  For freezing, again $w_i$ is 
distinct from the cosmological constant and from typical thawing values. 
The parameter $\beta$ shows tendencies toward the 
expected preference for small values for thawing and larger values for 
freezing, but the constraints are not tight.  Most importantly, the thawing 
and freezing classes can be distinguished from each other and from the 
cosmological constant. 
} 
\label{fig:th_fr_cont}
\end{figure*}

The probability distributions for the thawing and freezing fiducial 
cases are shown in Fig.~\ref{fig:th_fr_cont} (left and right panels 
respectively).  We find $w_i$ to be the parameter best constrained, 
with clear distinct regions of excluded parameter space.  In particular, 
the likelihood pulls away from the lower bound of the prior, preferring 
that $1+w_i>10^{-5}$ and thus demonstrating a distinction 
from the cosmological constant.  The flow deviation amplitude $\beta$ 
is less well constrained, but for the thawing fiducial case the likelihood 
has some preference for smaller values, in contrast to the freezing 
class.  For the freezing fiducial case, the contour prefers larger 
values of $\beta$, disfavoring a fit by thawing models.  As mentioned 
previously, the evolution rapidity $B$ is essentially unconstrained. 
We emphasize that the flow approach is not intended as a new parametrization 
to fit specific models, but rather a useful tool for distinguishing classes 
of physics, in particular with respect to the high redshift behavior.

For both thawing and freezing fiducial cases the 2-dimensional joint 
probability distributions for $(w_i, \beta)$ and $(w_i, B)$ are each almost 
decorrelated.  The exception is for large values of $\beta$ (large 
overall flow $F(a)$) where $w(a)$ need not start as negative since the 
evolution is pushed strongly toward $\Lam$CDM.  For $(w_i, B)$ the 
correlation is even weaker, and $w_i$ is insensitive to $B$ for most 
of its range except for very small values of $B$ (corresponding to an 
earlier flow deviation, evolving towards $\Lam$CDM), where $w_i$ can 
be further away from the cosmological constant behavior.  These 
results show that, while as expected a three parameter description 
of the equation of state cannot be tightly constrained, the key 
discriminating parameter of $w_i$ can be constrained.  This allows 
distinction (for this 1\% deviating fiducial at least) between each 
of the main classes of dark energy: a static cosmological constant, 
an initially thawing field, and an initially freezing field.

\section{Comparing Approaches \label{sec:compare}} 

Throughout most of this article we propounded the use of the flow 
quantity $F$ as a natural and slowly-evolving function, which could be 
relied upon to be approximately constant over a large number of 
$e$-folds.  Unlike the case of parametrization of the scalar field 
potential, for which we need to assume some knowledge of its form and 
hence restriction to a particular case, the flow function $F(a)$ 
possesses natural characteristics that allow us some insight on the 
dynamics of the field, on the sole assumption of matter domination, 
without having to assume some explicit form for $V(\phi)$. 

It is interesting to explore this argument further by showing that the 
opposite approach, a Taylor expansion of the potential as is often 
used (cf.\ \cite{liddlesahlen,hutpeir}), 
can lead to results which don't fully or faithfully describe the 
available and viable dynamics.  Indeed its consistency as a description 
of the potential can break down, as the quantities in which it is being 
expanded are not generally small.  Such is the case of the (not 
necessarily slowly rolling) slow-roll 
parameters $\epsilon(\phi)$ and $\eta(\phi)$, defined in terms of 
first and second derivatives of the potential, or the field displacement 
$\Delta \phi$.

The usual parametrization of the potential based on the slow-roll 
assumption takes 
\beq
V(\phi)/V(\phi_0=0) \sim 1+ V_1 \phi+ V_2 \phi^2 +...  \label{tayV}
\eeq 
where $\phi_0$ (taken to be zero without loss of generality) is the 
field value at some initial scale factor $a_i$, $V_0$ is the initial 
value of the potential energy density, and the coefficients of the 
expansion are expressed in terms of the usual slow-roll parameters 
\cite{liddle_parsons_barrow, hoffman_turner, kinney, easther_kinney, liddle_flow, peiris_easther} with $V_1= - \sqrt{16\pi \epsilon_i}$, and $V_2= 4\pi \eta_i$.

For the validity of Eq.~(\ref{tayV}) as a Taylor expansion, 
one needs either to rely on a 
sufficiently flat potential, by having small slow-roll parameters 
initially, or to take the expansion to be valid over a narrow region of 
the field trajectory, such that for the range of potential we are 
interested in (i.e.\ its evolution up to the present) $\Delta \phi$ is 
small.  In particular, we would want $\Delta\phi\ll(16\pi 
\epsilon)^{-1/2}$ and $\Delta\phi\ll(\epsilon/[\pi\eta^2])^{1/2}$ 
to have convergence, i.e.\ the linear being smaller than the constant 
term, and the quadratic term smaller than the linear term, respectively. 

However, remember there is no guarantee the potential is sufficiently 
flat, today or at high redshift. This in turn gives no grounds for 
taking small bounds on the slow-roll parameters, in Monte Carlo 
simulations or other analyses.  Indeed the detailed analysis of 
\cite{hutpeir} finds values of 
$\epsilon_i \approx 1$ and $\eta_i\approx5$, or greater, to be within 
the 68\% confidence limits contour at $z_i=3$ (for current distance 
data taken from supernovae, baryon acoustic oscillation, cosmic 
microwave background, and Hubble constant measurements; note they 
use the potential as a quadratic or cubic form, not a Taylor expansion 
per se).  If instead 
one tries to ensure validity of a (truly) slow-roll expansion by artificially 
imposing small $\epsilon_i$ (or equivalently starting the field at rest), 
one obtains strongly restricted dynamics as seen in the clear analysis 
and Fig.~9 of \cite{hutpeir}, which shows much narrower $w(a)$ evolutions 
than in the absence of this imposition. 

To highlight the influence of matter domination on the slow-roll parameters 
in the potential model, we tested starting the initial conditions at 
different redshifts and following the evolution of $\epsilon$ and $\eta$.  
If instead of imposing $\epsilon_i=1$, $\eta_i=5$ at $z_i=3$, we start with 
these values at higher redshift, $z_i=9$, then by $z=3$ the first slow-roll 
quantity has dropped to $\epsilon\approx0.01$ ($\eta$ stays fairly 
constant), clearly demonstrating the governing influence of matter domination. 

This reduction is insensitive to the initial redshift -- as long as it is 
at least one $e$-fold before the evaluation point at $z=3$. Thus, if one 
wants to carry out Monte Carlo simulations, either one can use a wide range 
for the parameters -- but the initial conditions should be set at least a 
few $e$-folds into matter domination, or for a more recent start one should 
use a restricted range of initial conditions as given by the matter 
dominated physics. 

In addition, in order to apply the Taylor expanded potential formalism one 
needs to rely on a small field excursion during the relevant dark 
energy evolution.  In terms of the slow-roll parameters, for 
$\epsilon\approx1$, $\eta\approx5$ the convergence conditions mentioned 
above require $\Delta\phi\ll0.14$ and $\Delta\phi\ll0.11$. 
With the aim of testing this assumption we compute 
the field displacement $\Delta \phi$ incurred for the flow models we 
considered previously, between the initial redshift and today, 
conservatively selecting those which most resemble $\Lam$CDM, since 
$\Delta \phi$ is smallest for these.  We will find that we cannot 
guarantee the necessary level of smallness to validate the potential 
approximation.  

From the expression for the scalar field kinetic energy (see above 
Eq.~\ref{eq:dphik}), the effective field excursion $\Delta \phi$ 
between the initial redshift, $z_i=9$ and today is 
\beq 
\Delta\phi = \int_{a_i}^1 \frac{da}{a}\, \sqrt{ 3\Omega_\phi \,(1+w) } \,. 
\eeq 
We compute $\Delta \phi$ for the flow models from Sec.~\ref{sec:condyn} 
that pass the 1\% distance criteria.  For thawing models the main 
contribution to $\Delta \phi$ is during the last $e$-fold or so, and 
accordingly we find a relatively small field displacement 
for thawers, of $\Delta \phi \approx 0.06$ for $w_i=-1+10^{-5}$ and up 
to $\Delta \phi \approx 0.4$ for the top of the range $w_i=-0.99$. 

On the other hand for the freezing behavior there is the possibility of 
a more substantial contribution to the field excursion throughout the 
evolution since the dark energy density is larger at early times and 
the equation of state finds its way towards $\Lam$ from larger $w_i$. 
For freezers we obtain $0.014 < \Delta \phi< 0.53$ between the smallest 
to the largest $w_i$, $-0.99<w_i<-0.36$. 

Both for thawers and freezers the leading influence on $\Delta \phi$ 
among the flow parameters is $w_i$.  
For thawers this is straightforward since $w(a)$ increases constantly 
with $w_i$ -- apart from those models which cross $F=1/3$ and turn 
around, receding back toward $\Lam$CDM: for those 
$\Delta\phi$ will be smaller than in the purely thawing case. 

The freezing dark energy case is different: large $w_i$ will cause the 
field to travel longer initially, before it slows down towards 
$w \approx -1$ in order to meet the distance criteria. For these fields 
particularly, it is crucial to initialize the evolution equations at 
high enough redshift, where the field is evolving the most and 
determining the conditions for its later dynamics.  Starting the 
evolution at $z=3$, only $1.4$ $e$-folds before the present, is 
problematic and can change the results.  For example, using 
$z_i=3$ Ref.~\cite{hutpeir} finds an 
average of $\Delta \phi=0.09 \pm 0.03$, out of their posterior 
distribution obtained from current data, 74\% of the models being 
freezer models.  
Our results using $z_i=9$ for the flow model give significantly 
different results; even for models that closely resemble $\Lam$ we find 
values as large as $\Delta \phi=0.34$ between $z=3$ and 
$z=0$, for freezer models.   Interestingly, in the absence of the slow-roll 
assumption either in the form of small slope or small field displacement, 
one finds the allowed field dynamics to be both richer and simultaneously 
able to remain close to $\Lam$CDM. 

This argument becomes all the more important if we remember we don't 
have any guarantee today of distance measurements agreeing with 
$\Lam$CDM to 1\%.  Relaxing this would further release the behavior of 
freezer models early on, and can give large field excursions, for 
example $\Delta\phi=0.87$ for an $n=1$ SUGRA model \cite{sugra}, 
with a maximum 3\% distance deviation for $z<1.7$. 

Thus the flow parameter approach appears to have a certain amount of natural 
motivation, from the long era of matter domination, and can offer 
a self-consistent approach.  It is also important to note that 
the appropriate initial conditions must be carefully considered for a 
Monte Carlo analysis of the field evolution. 

Because the relation $w'(w)$ can be written either in terms of the 
flow $F$ or the potential parameters, e.g.\ $\lambda\equiv -(1/V)dV/d\phi$ 
and $\Omega_\phi$, interchangeably through $F=(1+w)/[\Omega_\phi \lambda^2]$, 
the near constancy of $F$ implies that 
one cannot arbitrarily choose $\lambda$, $\Omega_\phi$, and $w_i$ 
in the Monte Carlo initial conditions during matter domination.  Moreover, 
if any of $1+w$, $\Omega_\phi$, or $\lambda^2$ start too large (including 
taking a uniform prior out to a relatively large upper bound), then 
generating realizations of the thawing class of models will be strongly 
suppressed.  As mentioned previously, even a large initial value of 
$\epsilon$ ($\sim\lambda^2$), if the evolution takes into account 
matter domination, becomes 
a much more restricted value by $z=3$ (this accords with Figure 9 of 
\cite{hutpeir} where once they restrict the initial value of $\epsilon$, 
a substantial fraction of randomly generated models become otherwise 
scarce thawers).  
While the flow approach is not without flaw, it does 
present an interesting alternative, one that innately incorporates the 
role of the long matter dominated era and the oldness of dark energy.

\section{Conclusions \label{sec:concl}} 

Dark energy does not exist in a vacuum.  
For dozens of $e$-folds of Hubble expansion 
it was dominated by matter and radiation, and has risen above them for 
only a fraction of the last $e$-fold.  This has a definite effect on the 
dynamical evolution of dark energy, aging or mellowing it.  In particular, 
a wide range of models exhibit a nearly constant flow parameter, relating 
the deviation of the equation of state from $-1$, the deviation of the dark 
energy density from 0, and the slope of the potential.  It is not until 
$z\gtrsim2$ that the flow parameter, and the equivalent relation between 
$w'$ and $w$, is generally free to vary. 

This characteristic suggests the flow parameter is an interesting quantity 
to use for parametrizing the dark energy evolution, one that can 
intrinsically take into account the oldness of dark energy.  We have 
examined an evolutionary model for the deviations of the flow parameter 
from constancy as matter domination wanes and acceleration occurs. 
This gives rise to a variety of dynamical behaviors in the $w'$-$w$ 
phase space, incorporating thawing, freezing, and ``refreezing'' classes. 
Observational constraints on the dark energy equation of state, e.g.\ through 
distance-redshift measurements, bound specific regions in the $w'$-$w$ 
plane.  For example, scanning 250,000 models, we illustrate the region, 
and its component 
classes, of maximum deviation allowed away from $w=-1$.  We also identify 
an interesting ``waist'' phenomenon -- independent of the standard 
$w_0$-$w_a$ parametrization -- where at a certain redshift the equation 
of state is tightly constrained. 

With an MCMC analysis one can quantify the bounds on the values of the 
flow parameters allowed by distance measurements.  In particular 
this permits determination of the high redshift dark energy 
equation of state, a key parameter for the important physics question of 
distinguishing between a cosmological constant, initially thawing models, 
and initially freezing models.  Finally, we emphasize that taking into 
account the maturity of dark energy is important for MCMC generation of 
models since random initial conditions, e.g.\ on the potential, do not 
appropriately sample the ``diversity under constraint'' of old dark energy.

\acknowledgments 
We thank Sarah Bridle, Dragan Huterer, Andrew Liddle, Hiranya Peiris, 
and An{\v{z}}e Slosar for discussions, 
and are especially grateful to Martin Kunz and David Parkinson for helpful 
advice. MC thanks the Institute for the Early Universe at Ewha University, IPMU in Japan, and Shinji Tsujikawa and the Tokyo University of Science, for 
hospitality. We acknowledge the use of the computer 
system at the Astronomy Centre at the University of Sussex.  This work has 
been supported in part by the Director, Office of Science, Office of High 
Energy Physics, of the U.S.\ Department of Energy under 
Contract No.\ DE-AC02-05CH11231.

\end{document}